# Superconductivity and Magnetism in a New Class of Heavy-Fermion Materials

J. D. Thompson[a], R. Movshovich[a], Z. Fisk[a,b], F. Bouquet[c], N. J. Curro[a], R. A. Fisher[c], P.C. Hammel[a], H. Hegger[a,d], M. F. Hundley[a], M. Jaime[a], P.G. Pagliuso[a], C. Petrovic[a,b], N. E. Phillips[c], and J. L. Sarrao[a]

[a]Los Alamos National Laboratory, Los Alamos, NM 87545 USA
[b]NHMFL/Florida State University, Tallahassee, FL 32306 USA
[c]Lawrence Berkeley National Laboratory, University of California, Berkeley, CA 94270 USA
[d]Bayer AG, Leverkusen, Germany

*Abstract:*
We report a new family of Ce-based heavy-fermion compounds whose electronic specific heat coefficients range from about 400 to over 700 mJ/mole Ce-K$^2$. Crystals in this family form as Ce$_n$T$_m$In$_{3n+2m}$, where T= Rh or Ir, n=1 or 2, and m=1, with a tetragonal structure that can be viewed as n-layers of CeIn$_3$ units stacked sequentially along the c-axis with intervening m-layers of TIn$_2$. Ambient and high pressure studies show that the quasi-2D layers of CeIn$_3$ produce unconventional superconducting and magnetic ground states. This family should enable new understanding of the relationship between magnetism and superconductivity in heavy-fermion materials and more generally of why heavy-fermion superconductivity prefers to develop in one structure type and not another.

*Corresponding author:*
Joe D. Thompson
MS K765
Los Alamos National Laboratory
Los Alamos, NM 87545 USA
Fax: (505) 665-7652
Email: jdt@mst.lanl.gov

There is ample evidence that the superconductivity and small-moment magnetism found in cerium- and uranium-based heavy-fermion materials are unconventional and that the physics of these ground states may be related.[1] In particular, it appears that the unconventional superconducting ground state always appears in proximity to the equally unconventional magnetic state. This has led to the logical assumption that some limit of the same mechanism is responsible for both. Recent discoveries [2] of a pressure-induced transition from magnetic to superconducting states in the same material hold promise for more detailed understanding of the relationship between these two states. However, in several of these examples, the pressure required to induce the transition is rather high, making it particularly challenging to make more than just the most basic measurements, such as electrical resistivity. Ideally, what is needed to make progress on the heavy-fermion problem is a single family of materials whose ground states can be tuned easily with modest pressure and/or chemical substitutions. The $ThCr_2Si_2$ structure type is one such class of materials in which heavy-fermion magnetism and superconductivity seem to be favored, but we do not know why this is the case. The first known heavy-fermion superconductor $CeCu_2Si_2$ is a member of this class, and, by changing the precise Ce/Cu ratio in the compound, it can be tuned at ambient pressure between superconducting and magnetic states and for a fixed ratio from non-superconducting to superconducting states with pressure.[3] However, its delicate crystal chemistry has made it a very challenging system to study. The same is true of its relative $CeNi_2Ge_2$. Even if there were no chemistry difficulties, it would be valuable to have another Ce-based materials type that would allow us to understand more broadly the relationship between magnetism and superconductivity and even more generally why heavy-fermion superconductivity prefers to develop in one structure type and not another. Here, we report on a new family of materials that holds promise for making progress in these regards.

These new materials form with chemical compositions $Ce_nT_mIn_{3n+2m}$, where T= Rh or Ir. All members with m=1, n=1 or 2 grow readily out of an In-rich flux as single crystals with characteristic size 1 cm x 1 cm x several mm. Powder x-ray patterns obtained on crushed single crystals show [4] that the compounds grow with a tetragonal unit cell that can be viewed as n-layers of $CeIn_3$ units stacked sequentially along the c-axis with intervening m-layers of $TIn_2$. Except for Eu, single-layer (n=1) and bi-layer (n=2) variants also grow with the light rare earths La through Gd.[5] The La-based materials are, as expected, Pauli paramagnets to 50 mK. Lattice parameters for the Ce-based compounds are given in Table 1. The in-plane lattice parameter $a_0$ is a measure of the Ce-Ce spacing within the cubic $CeIn_3$ units, and the c-axis parameter reflects the Ce-Ce spacing perpendicular to the planes for n=1. For n=2, the c-axis parameter is the repeat distance of $CeIn_3$ bilayers and $a_0$ is essentially the nearest neighbor distance between Ce atoms in adjacent layers. As might be expected from the quasi-2D structure, the magnetic susceptibility of these materials also is anisotropic and depends on the value of n as well as the transition metal T. In all cases, the effective magnetic moment, obtained from plots of the inverse susceptibility versus temperature above 200 K, is reduced somewhat from the Hund's rule value of 2.54 $\mu_B$ for $Ce^{3+}$, indicating the presence of crystal-field effects, and the low temperature susceptibility is always larger when a magnetic field is applied parallel to the c-axis. The ratio of $\chi_c/\chi_a$ at low temperatures is larger for n=1 than for n=2 and for T=Rh than T=Ir, ranging from greater than 2 for n=1,T=Rh to 1.2 for n=2,T=Ir. The quality and degree of chemical order in these materials are reflected in part by values of the resistivity ratio ρ(300 K)/ρ(4 K) that are on the order of 50-100 and by the appearance of very narrow lines in NQR spectra. Additional details of the structural and physical properties will be given elsewhere; in the following, we briefly discuss some of the low temperature properties of the n=1 compounds and mention those of the less-studied n=2 family.

In Fig.1 we plot the temperature dependence of the resistivity ρ, static susceptibility χ, and the specific heat divided by temperature C/T for $CeRhIn_5$. All exhibit a feature at 3.8 K that is associated with magnetic order. Approximately 30% of Rln2 entropy is found below 3.8 K, suggesting substantially reduced moment ordering in a crystal-field doublet ground state. The remaining 70% Rln2 magnetic entropy is recovered on warming to 18-20 K. It is somewhat difficult to define a Sommerfeld coefficient from the C/T data just above $T_N$, but a simple entropy-balance construction, $S(T_N- \varepsilon)=S(T_N+ \varepsilon)$, gives γ ≈ 400 mJ/mole-$K^2$. The susceptibility reaches a maximum near $T_{\chi m}$=7.5 K before dropping more steeply at $T_N$. Cerium-based correlated electron materials, in which J=5/2, commonly exhibit a low temperature maximum in their susceptibility that is expected from the theory of orbitally degenerate Kondo impurities.[6] The specific heat and crystal structure of $CeRhIn_5$, however, suggest that its ground state is doubly degenerate, in which case the Kondo effect produces the maximum susceptibility at T=0. An alternative interpretation for the maximum comes from the quasi-2D structure. It is known [7] that a 2D, spin-1/2 Heisenberg system exhibits a maximum susceptibility at $T_{\chi m}$≈ 0.93 |J|, which for $T_{\chi m}$=7.5 K, gives |J|=8 K. Any deviation from purely 2D exchange produces long range order at $T_N$ ≈S(S+1)|J|/2= 3K, which is close to the experimentally observed value of 3.8 K. This simple picture, which neglects Kondo effects that are undoubtedly present, may at least qualitatively account for the susceptibility maximum in $CeRhIn_5$.

The qualitative view of the magnetic state inferred from specific heat and susceptibility measurements is consistent with more microscopic information obtained from $^{115}$In NQR data. In the $CeRhIn_5$ structure, there are two inequivalent In sites: a single In(1) site, analogous to the single In site in $CeIn_3$, and four In(2) sites, two on each of the lateral faces of the unit cell that are equidistant above and below the Rh layer. Results of an analysis of NQR measurements on the In(1) site are shown in Fig.2. The internal magnetic field produced by magnetic order grows rapidly below $T_N$ and saturates to a value of 0.17 T. This internal field compares to 0.5 T found at the equivalent In site in bulk $CeIn_3$, for which neutron diffraction studies show an ordered moment of 0.4 $\mu_B$.[8] Assuming a similar hyperfine coupling in both materials, the ordered moment in $CeRhIn_5$ should be 0.1-0.2 $\mu_B$ and, by symmetry considerations, lies in the tetragonal plane. A fit of $H_{int}$ versus $(1-T/T_N)^\beta$ near $T_N$ gives a critical exponent β=0.25±0.03, which is about one half the mean field value of β=0.5. Similarly rapid growth of the sublattice magnetization below $T_N$ also has been found in $La_2CuO_4$ in which a reduced critical exponent is related to its 2D magnetism.[9] Preliminary analysis of NQR spectra from the In(2) sites suggest that the ordered moments maintain a constant magnitude within each plane, but their ordered direction is modulated by a spiral rotation along the c-axis that is incommensurate with the lattice.

Application of pressure to the heavy-fermion antiferromagnet $CeIn_3$ drives its Néel temperature smoothly from 10 K at atmospheric pressure toward zero at a critical pressure of 25 kbar, where superconductivity develops with $T_c$ ≈ 0.25K.[10] The pressure response of $CeRhIn_5$ is qualitatively different. From resistivity measurements, we derive [11] the T-P phase diagram shown in Fig. 3b. The resistivity feature at 3.8 K shown in Fig.1 moves slowly with applied pressure to higher temperatures at a rate of about 9mK/kbar. This feature is present at 14.5 kbar but not at higher pressures. At 16.3 kbar, there is a broad transition beginning near 2 K to a zero resistance state; the transition width sharpens with increasing pressure to less than 0.05 K at 21 kbar, where the onset of superconductivity is at 2.17 K. The abrupt loss of a signature for $T_N$, the sudden appearance of superconductivity, and rapid sharpening of the transition width suggest a first-order like transition at a critical pressure between 14.5 and 16.3 kbar. AC susceptibility measurements on a second crystal reproduce the appearance of superconductivity and show a perfect diamagnetic response below $T_c$, which is almost an order of magnitude higher than in bulk $CeIn_3$.

Results of specific heat measurements on $CeRhIn_5$ at 19 kbar are shown in the upper panel of Fig. 3. There are several points to note about these data. C/T begins to increase more rapidly below about 5 K, where C/T = 390 mJ/mole-$K^2$, reaches a plateau near 2.5 K, and then shows a well-defined feature at $T_c$. Below $T_c$, C/T is linear in temperature, i.e., $C \propto T^2$ and not exponential as expected for BCS superconductivity. This power-law dependence of C is expected for a superconductor in which there are line-nodes in the gap function.[12] The state reflected in the plateau of C/T near 2.5 K also is found as a weak feature in resistivity measurements, and its pressure dependence is shown as the transition labeled $T_?$ in Fig. 3b. This phase transition first appears in resistivity measurements at pressures near 10 kbar, well below the pressure necessary to induce superconductivity, and persists in the pressure regime of superconductivity. We do not know its origin but believe it involves an instability of the Fermi surface. In this regard, we observe that the jump in specific heat $\Delta C$, measured from the value of C/T ≡ γ = 390 mJ/mole-$K^2$ at 5 K to the maximum at $T_c$, gives $\Delta C(T_c)/\gamma T_c$=1.5±0.1, which is close to the weak coupling value of 1.43. That is, the transition $T_?$ plus the superconducting transition appear to gap the entire Fermi surface. Further, the $T_?$ transition is unchanged in a field of 9 T; whereas, in this field, superconductivity is suppressed below 0.3 K. We are led to the conclusion that the $T_?$ transition is one to a charge- or spin-density wave state that gaps part of the Fermi surface and that this density wave state coexists with superconductivity. What relationship there might be between the density wave and Néel states remains to be explored.

We turn now to the other n=1 member of this family, $CeIrIn_5$. Its thermodynamic and transport properties at low temperatures are summarized in Fig. 4. At 0.38 K, there is a diamagnetic transition in ac susceptibility that is coincident with a jump in heat capacity, providing clear evidence of bulk superconductivity. By comparing the magnitude of the ac susceptibility response of a sample of tin with similar size and shape at its $T_c$ to that of $CeIrIn_5$ at its superconducting transition temperature, we estimate that the response below 0.38 K shown in Fig. 4 corresponds to perfect diamagnetism. Just above $T_c$, C/T is essentially constant and gives a Sommerfeld coefficient γ= 720 mJ/mole-$K^2$. From the average of measurements on three different crystals, the specific heat jump $\Delta C$ at $T_c$ is equal to $(0.76 \pm 0.05)\gamma T_c$. This ratio $\Delta C(T_c)/\gamma T_c$ is comparable to that found in other heavy-fermion materials, eg. $UPt_3$ [13], and is smaller than the value for $CeRhIn_5$ at 19 kbar, but in that case, we also included the effect of the $T_?$ transition. The specific heat data of $CeIrIn_5$ below $T_c$ are well-fit to the sums of nuclear-Schottky, $T^2$ and T-linear contributions. (A nuclear Schottky term is expected due to the large nuclear quadrupole moment of In and presumably is present in the superconducting state of $CeRhIn_5$, but at temperatures lower than shown in Fig. 3.) A significant $C \propto T^2$ contribution suggests that, as in $CeRhIn_5$, the gap function has line-nodes. The large T-linear term of $200 \pm 50$ mJ/mole-$K^2$ indicates the presence of ungapped quasiparticle states below $T_c$ and provides further support for the existence of zeros in the gap. The temperature dependence of the thermal conductivity, which is insensitive to the nuclear Schottky, also is described well [14] from $T_c$ to 50 mK by the sum of linear and quadratic terms that are consistent with corresponding terms in the specific heat.

A very peculiar aspect of the data shown in Fig. 4 is that the resistivity drops to zero at $T_0$=1.2 K, or at least to less than 1% of the resistivity above 1.2 K, without an obvious thermodynamic signature. Our original belief was that this resistive transition was extrinsic. However, additional measurements suggested that this is intrinsic. Measurements [15] of the specific heat, ac susceptibility and electrical resistivity, with all three measurements made on each of three separately grown crystals, in magnetic fields to 9 T applied parallel and perpendicular to each sample's c-axis show that the anisotropic response of $T_c$ determined by specific heat and ac susceptibility is identical to that determined resistively. That is, anisotropy in the field dependences of $T_c$ and $T_0$ is identical. This would seem to imply that both transitions arise from a common underlying electronic structure. Preliminary studies of the response of $CeIrIn_5$ to pressure and to Rh-doping also provide evidence that the thermodynamic and resistive transitions are intrinsic and intimately linked: in both cases the bulk $T_c$ increases and approaches $T_0$, which is relatively insensitive to perturbation. Presently, we have no definitive explanation for the origin of the resistive transition, but the data suggest that it is intrinsic to $CeIrIn_5$, arising possibly from filaments of locally phase coherent electron pairs for $T_c<T<T_0$ that become globally coherent through out the sample at $T_c$..

The n=2 variants of these materials also appear to be quite interesting, and their study in parallel with the n=1 compounds should provide insight into the role of spatial dimensionality in controlling the nature of their heavy-fermion ground states. Interestingly, the Sommerfeld specific heat coefficient of the n=2 members is the same, to within 10%, of that for the corresponding n=1, T= Rh and Ir materials, which implies that the mechanism responsible for producing the heavy-mass state depends on the transition element T and not n. This further supports the generally held belief that the large $\gamma$, a hallmark of heavy-fermions, arises from a local many-body correlation of the f-electrons with the sea of ligand electrons. However, the ground state is affected by n. $Ce_2RhIn_8$ orders antiferromagnetically at 2.8 K, which is 1 K lower than in $CeRhIn_5$. Very nearly the same magnetic entropy is liberated below their respective magnetic transitions. On-going studies of the effect of pressure on $Ce_2RhIn_8$ show that the resistive signature for $T_N$ changes only slightly with pressure and disappears abruptly, but at a lower pressure (less than 7 kbar) than in $CeRhIn_5$. At 17 kbar and higher, the positive slope of the resistivity increases below approximately 2 K and may signal the onset of another transition. While the ground state of $Ce_2RhIn_8$ appears similar to that of $CeRhIn_5$, this is not the case for the Ir-containing compounds. $Ce_2IrIn_8$ remains a heavy-fermion paramagnet to 50 mK, with no evidence for a phase transition. As shown in Fig. 5, there is an interesting progression in the temperature at which a broken symmetry state develops in this family of materials. When that temperature is plotted as a function of the Sommerfeld coefficient $\gamma_V$, normalized by the unit cell volume per Ce atom, magnetism appears at small $\gamma_V$ and superconductivity and paramagnetism at larger values. A similar correlation has been noted [16] previously among U-based heavy-fermion materials.

In summary, we have found a new family of heavy-fermion compounds whose ground state can be tuned easily by pressure and chemical substitution. All members of this family form as high quality single crystals with consistently reproducible properties, making them particularly amenable to careful, quantitative study. Unlike previous examples, these materials also allow investigation of the role of dimensionality in controlling the heavy-fermion state.

Acknowledgments: Work at Los Alamos was performed under the auspices of the U. S. Department of Energy. The work at Berkeley was supported by the Director, Office of Energy Research, Office of Basic Energy Sciences, Materials Sciences Division of the U. S. Department of Energy under contract DE-AC03-76SF00098. Z. F. acknowledges support through NSF grants DMR-9870034 and DMR-9971348.

Table 1. Lattice parameters, Néel or superconducting transition temperatures, and specific heat Sommerfeld coefficients for the $Ce_nT_mIn_{3n+2m}$ family of materials.

| Compound | $a_0$ (Å) | c (Å) | $T_N$ or $T_c$ (K) | γ (mJ/mole Ce-$K^2$) |
|---|---|---|---|---|
| $CeRhIn_5$ | 4.652(1) | 7.542(1) | 3.8 | ≈400 |
| $CeIrIn_5$ | 4.668(1) | 7.515(2) | 0.4 | 720 |
| $Ce_2RhIn_8$ | 4.665(1) | 12.244(5) | 2.8 | ≈400 |
| $Ce_2IrIn_8$ | 4.671(2) | 12.214(6) | -- | 700 |

Figure captions

Fig. 1. (a) Magnetic specific heat (C = $C_{total}$(CeRhIn$_5$ )-C(LaRhIn$_5$)) divided by temperature (circles), in-plane magnetic susceptibility $\chi$ (triangles) and temperature derivative of the electrical resistivity d$\rho$/dT (solid line) as a function of temperature for CeRhIn$_5$ at atmospheric pressure. (b) In-plane magnetic susceptibility (solid curve) and magnetic contribution to the electrical resistivity (dotted curve) of CeRhIn$_5$ over a broader temperature range. The resistivity of LaRhIn$_5$ was subtracted from the total resistivity of CeRhIn$_5$ to obtained the magnetic contribution.

Fig. 2. Temperature dependence of the internal field at the In(1) site that develops in CeRhIn$_5$ below $T_N$. The solid curve is a fit to $(1-T/T_N)^\beta$. See text for details.

Fig. 3. (a) Magnetic contribution to the specific heat C divided by temperature (open squares) and electrical resistivity (solid circles) for CeRhIn$_5$ at 19 kbar. The magnetic contribution to C was estimated by subtracting the specific heat of LaRhIn$_5$ at atmospheric pressure from the total specific heat of CeRhIn$_5$ at pressure. (b) The temperature-pressure phase diagram determined from resistivity measurements. $T_N$, $T_c$ and $T_?$ correspond to the Néel temperature, onset temperature for superconductivity, and the transition to an unknown state, respectively.

Fig. 4. Magnetic specific heat (C = $C_{total}$(CeIrIn$_5$ )-C(LaIrIn$_5$)) divided by temperature (circles), ac magnetic susceptibility $\chi$ (triangles) in arbitrary units and electrical resistivity $\rho$ (squares) as a function of temperature for CeIrIn$_5$. The solid line through C/T data is a fit as described in the text.

Fig. 5. Temperature at which a phase transition develops in the $Ce_nT_mIn_{3n+2m}$ family of materials. The horizontal axis is the Sommerfeld coefficient of specific heat normalized by the average volume of a Ce atom in a unit cell. Solid symbols denote magnetism, open symbols superconductivity or paramagnetism.

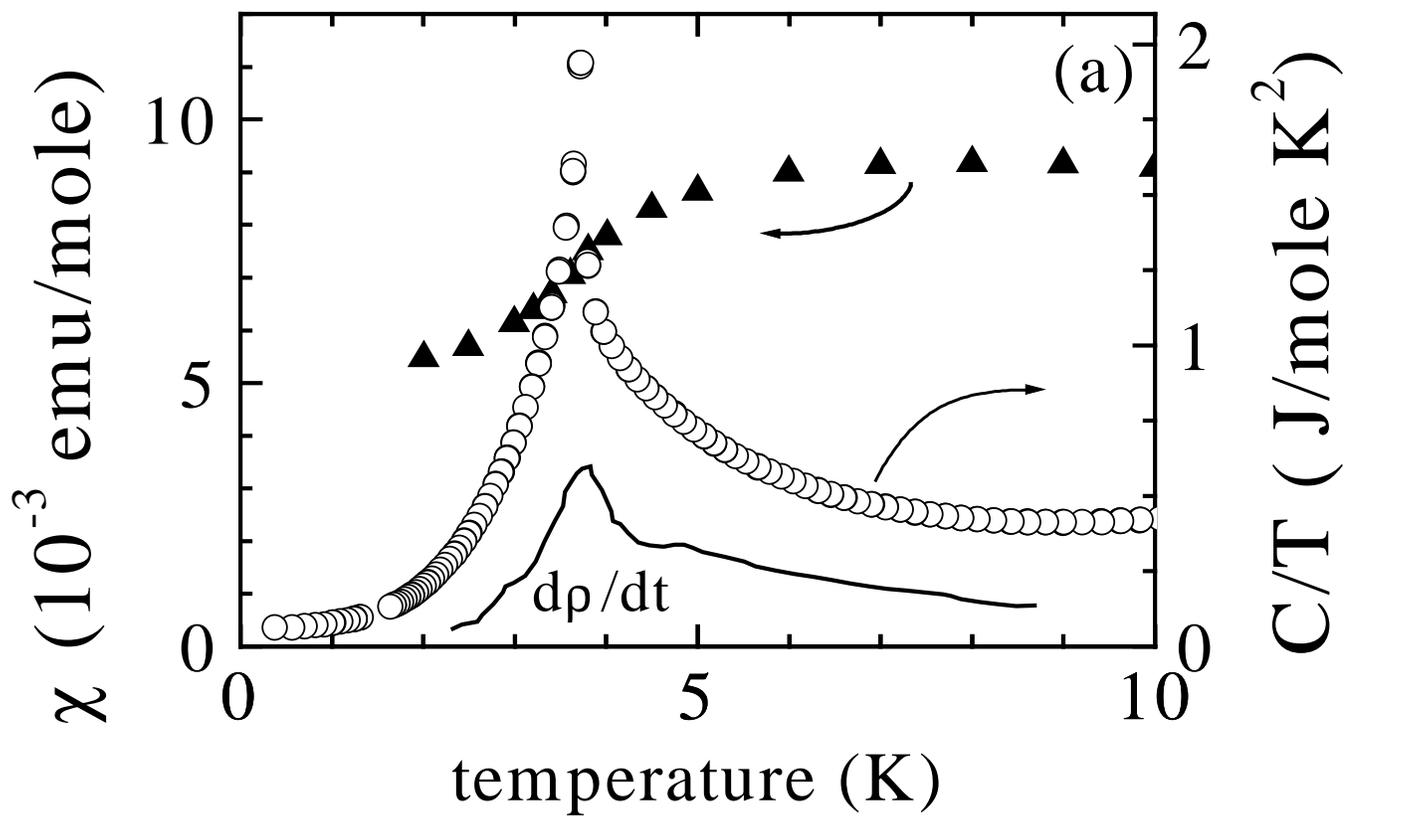

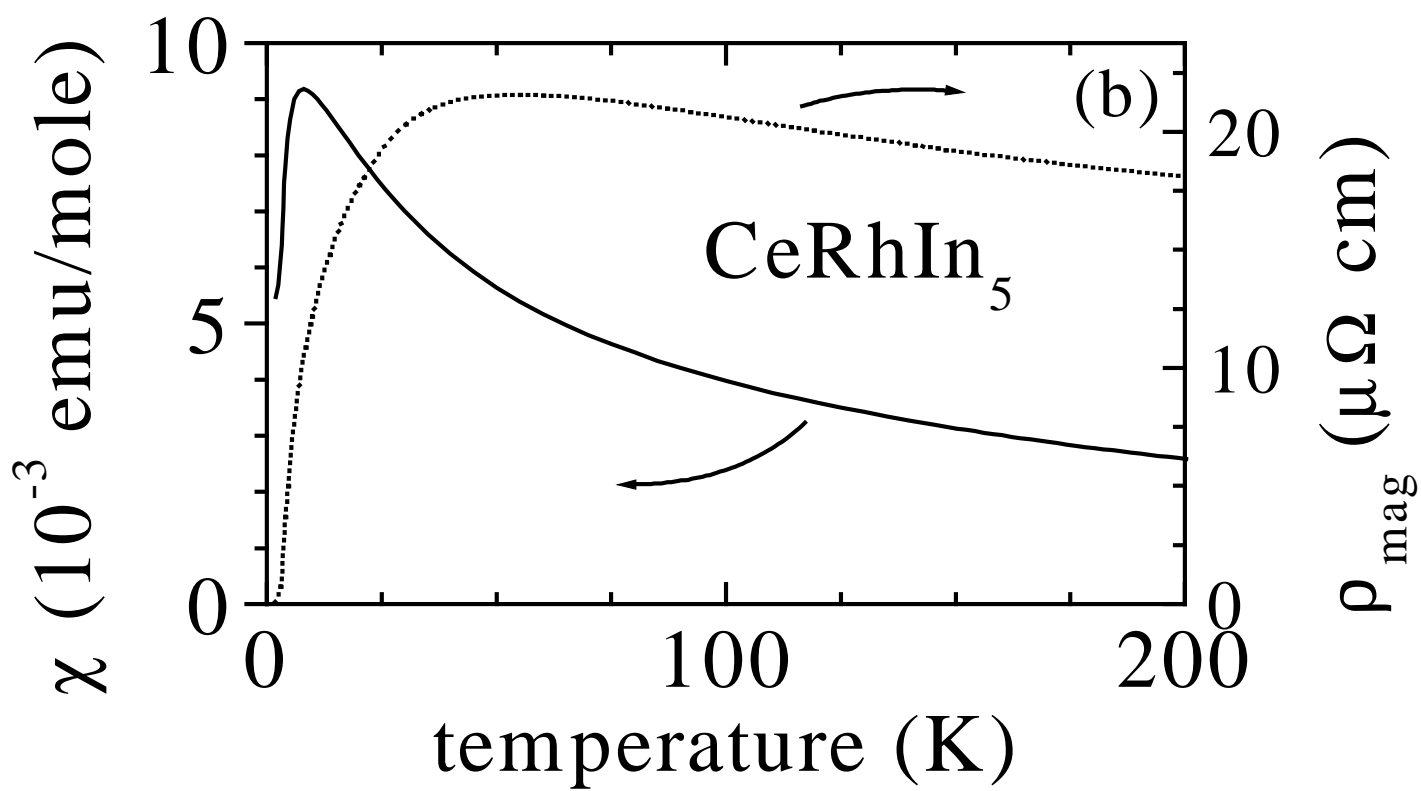

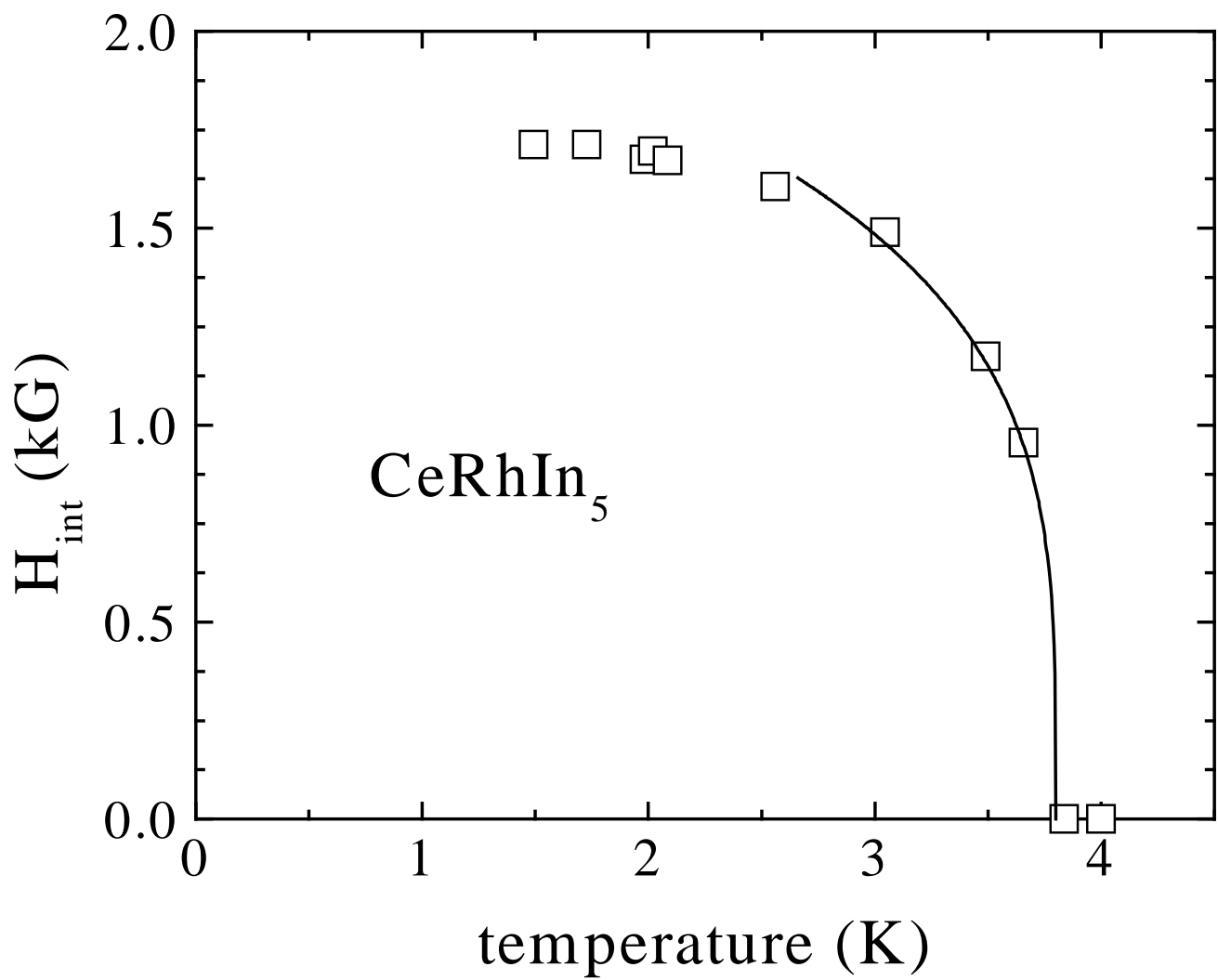

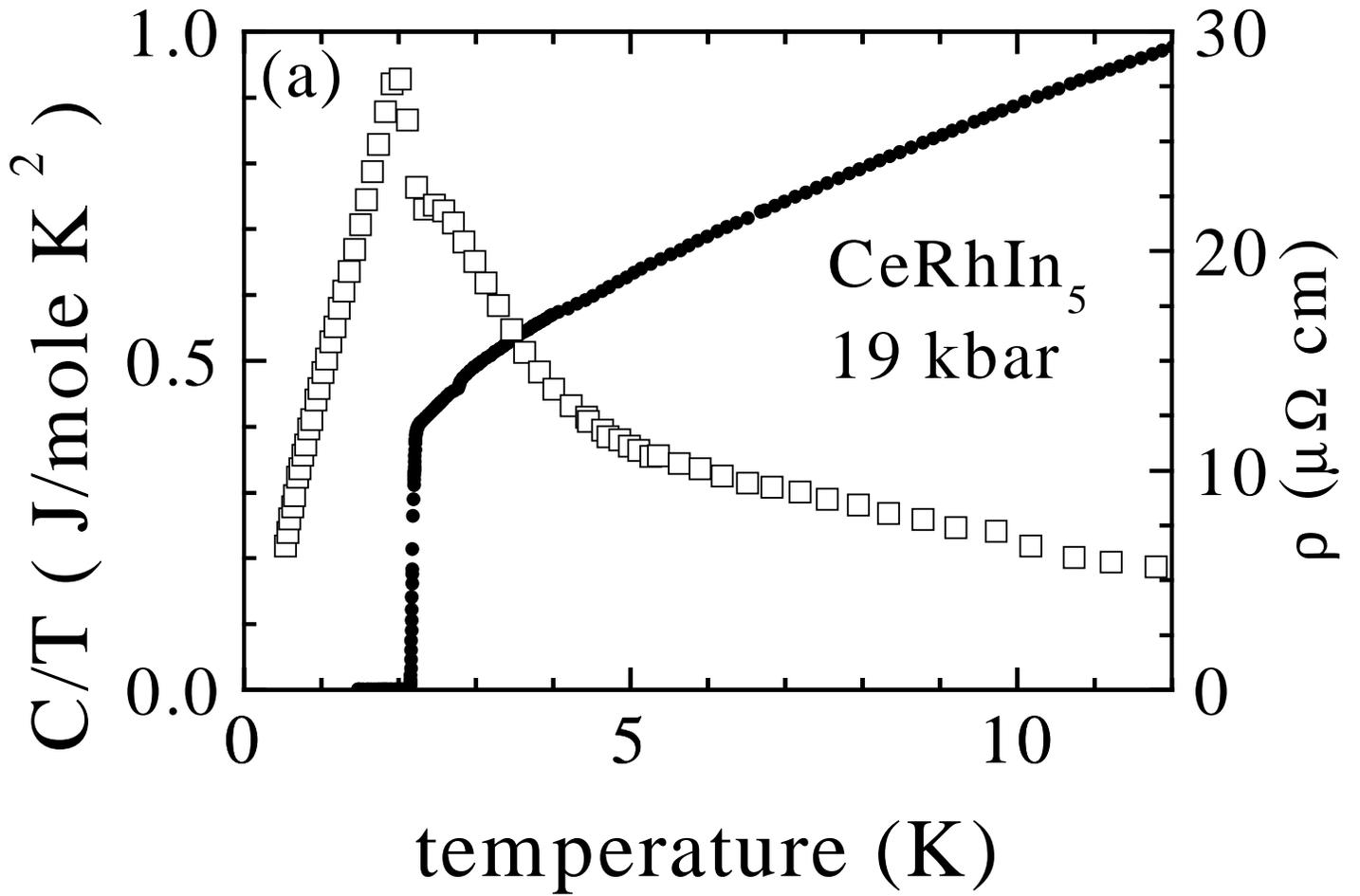

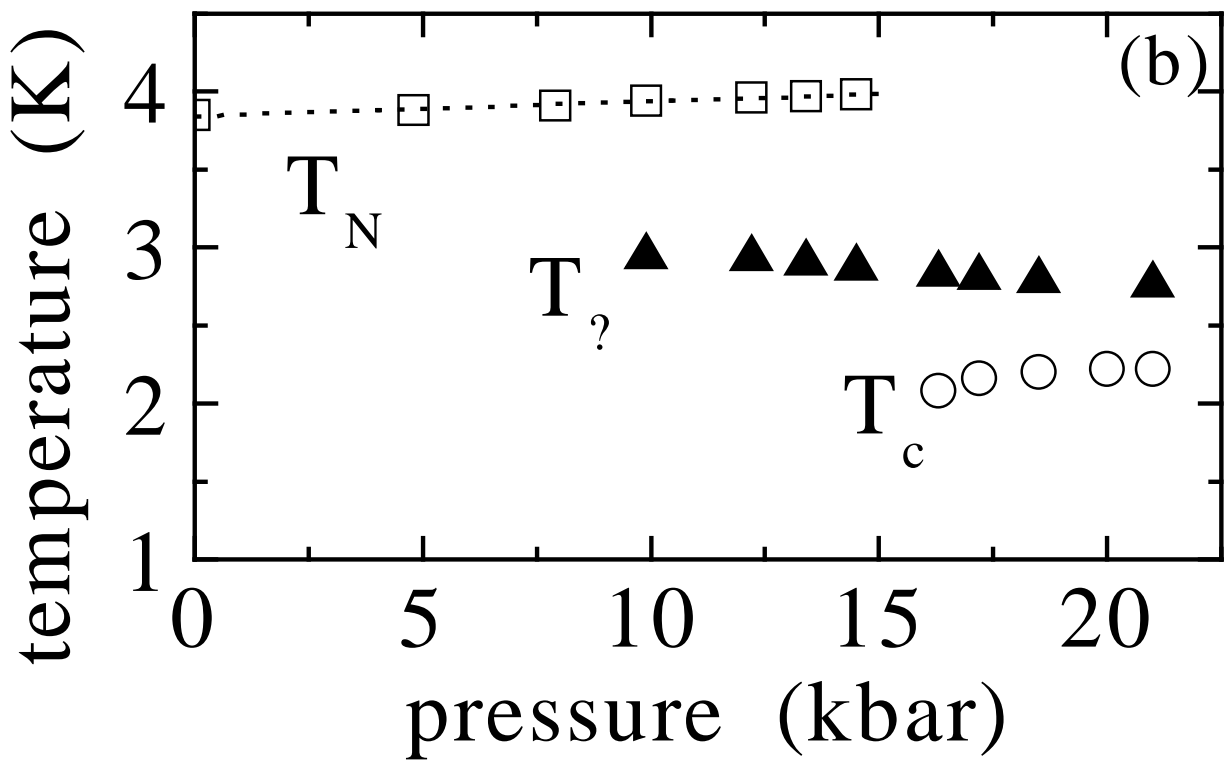

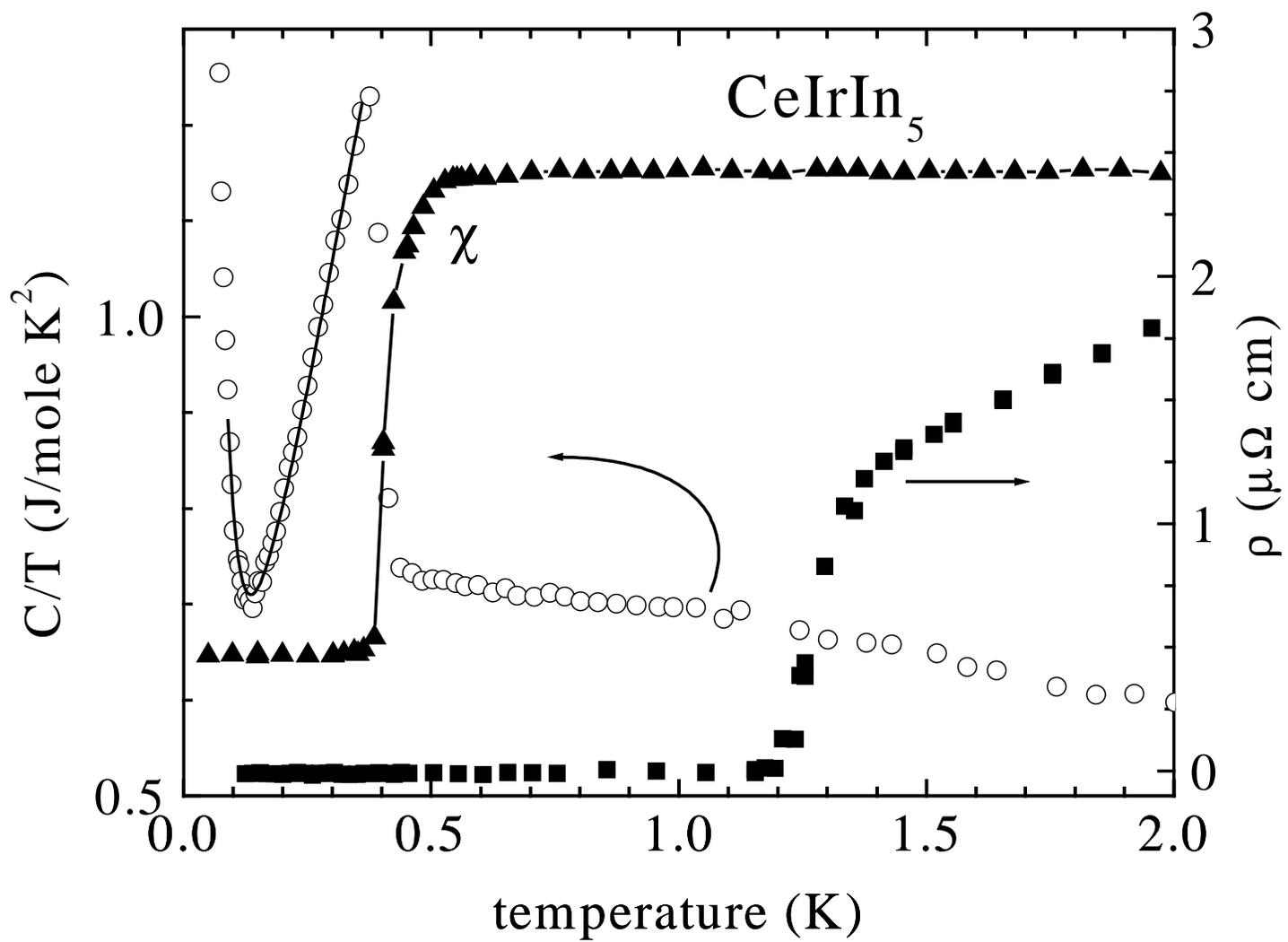

Thompson, *et al.* Fig. 5

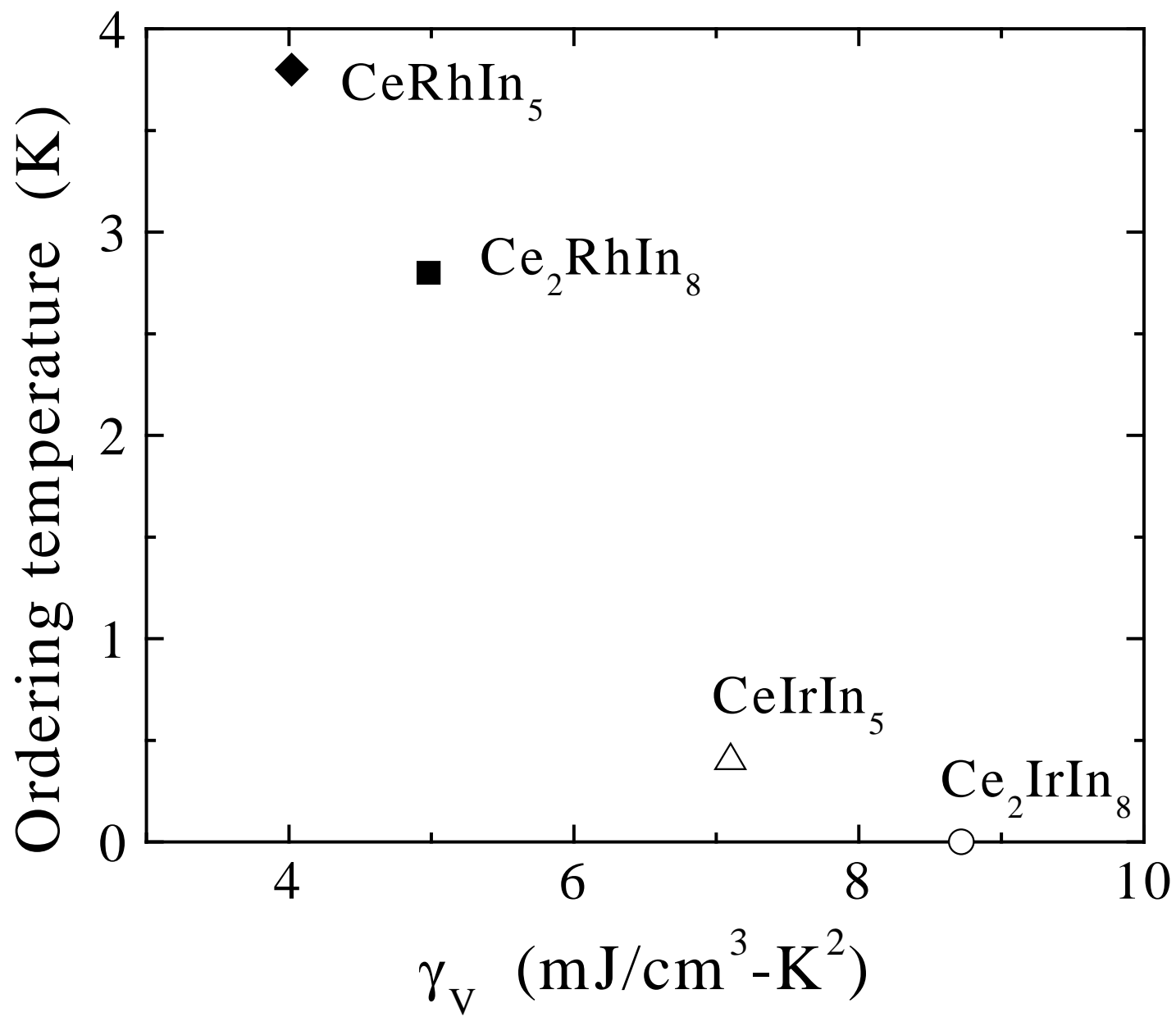